# Cross-sectional personal network analysis of adult smoking in rural areas

Bianca-Elena Mihăilă[1,2], Marian-Gabriel Hâncean[1,2], Matjaž Perc[3-6], Jürgen Lerner[7], Iulian Oană[1,2], Marius Geantă[2], José Luis Molina[8], Cosmina Cioroboiu[2]

[1] Department of Sociology, University of Bucharest, Panduri 90-92, 050663 Bucharest, Romania

[2] Center for Innovation in Medicine, Theodor Pallady Blv. 42J, 032266 Bucharest, Romania

[3] Faculty of Natural Sciences and Mathematics, University of Maribor, Koroška cesta 160, 2000 Maribor, Slovenia

[4] Community Healthcare Center Dr. Adolf Drolc Maribor, Vošnjakova ulica 2, 2000 Maribor, Slovenia

[5] Complexity Science Hub Vienna, Josefstädterstraße 39, 1080 Vienna, Austria

[6] Department of Physics, Kyung Hee University, 26 Kyungheedae-ro, Dongdaemun-gu, Seoul, Republic of Korea

[7] Department of Computer and Information Science, University of Konstanz, 78457 Konstanz, Germany

[8] Department of Social and Cultural Anthropology, Universitat Autònoma de Barcelona, 08193 Bellaterra, Barcelona, Spain

# Abstract

While research on adolescent smoking is extensive, little attention has been given to smoking behaviors among rural middle-aged and older adults. This study examines the role of personal networks and sociodemographic factors in predicting smoking status in a rural Romanian community. Using a link-tracing sampling method, we gathered data from 76 participants out of 83 in Lerești, Argeș County. Face-to-face interviews collected sociodemographic data and network information, including smoking status and relational dynamics. We applied multilevel logistic regression models to predict smoking behaviors (current smokers, former smokers, and non-smokers) based on individual characteristics and network influences. Results indicate that social networks significantly influence smoking behaviors. For current smokers, having a smoking family member greatly increased the odds of smoking (OR = 2.51, 95% CI: 1.62, 3.91, $p < 0.001$). Similarly, non-smoking family members increased the likelihood of being a non-smoker (OR = 1.64, 95% CI: 1.04, 2.61, $p < 0.05$). Women were less likely to smoke, highlighting sex differences in behavior. These findings emphasize the critical role of social networks in shaping smoking habits, advocating for targeted interventions in rural areas.

# 1. Introduction

Tobacco smoking remains the leading cause of preventable diseases [1] and the top contributor to premature deaths in the European Union (EU) [2]. EU measures include regulations like the Tobacco Products Directive (2014/40/EU) and campaigns such as *Ex-Smokers are Unstoppable* (2014-2016). The last EU informational campaign was in 2016, with subsequent evaluations recommending SMART goals for future campaigns [3]. Since then, the focus has shifted to national campaigns, with little known about their effectiveness.

The World Health Organization reports alarming trends, with tobacco killing over eight million people annually and affecting 1.3 million non-smokers through second-hand smoke [4]. Smokers face significantly higher risks of cardiovascular diseases, with mortality rates nearly tripling compared to non-smokers [5]. Smoking is also linked to diabetes [6], cataracts [7], gastrointestinal diseases [8], and various cancers, particularly lung cancer [9,10].

Smoking rates vary across EU regions: 28% in Eastern Europe and 20% in Northern Europe [11]. In 2019, Eurostat data revealed that 22.3% of men and 14.8% of women aged 15 and older smoked daily [12]. Additionally, 14% of individuals aged 15-24 were daily smokers [13]. Despite progress, strategies are insufficient to reduce smoking incidence by 30% by 2025 [14].

Most of the tobacco literature has largely overlooked adults and older people, focusing instead on younger populations. This highlights a critical gap, underscoring the urgent need for research that includes these demographics. Such studies are vital for comprehensively understanding smoking behaviors across all age groups. This ensures that interventions and policies are well-informed and effectively tailored to meet the needs of the entire population Additionally, rural populations have also been neglected despite evidence suggesting higher smoking rates in these areas [15]. The very few studies on rural communities emphasize smoking trends among females without focusing on gender-specific strategies [16]. We want to emphasize that most of these studies are very specific and refer to rural areas in the United States [15,17].

To address these gaps, our aim is to expand the understanding of smoking behaviors by studying often overlooked segments of the population in Romania, a representative country of Eastern Europe. Specifically, our study focuses on adults living in rural areas to bridge significant gaps identified in the existing body of literature. We investigate the relationship between social networks and smoking habits, examining how the intricate pattern of interpersonal ties within rural communities is associated with smoking behavior. This approach not only adheres to the necessity for inclusivity in research but also emphasizes the multifaceted interactions present in rural environments, particularly in the Eastern European context.

Romania has a high smoking prevalence: 30.6% of men and 7.5% of women smoke [18]. Among 15-24-year-olds, 10.2% smoke daily, and 9.8% occasionally [18]. Additionally, 71.5% of daily smokers light up within 30 minutes of waking [19]. Literature on Romanian smoking primarily focuses on adolescents. Researchers highlight the significant impact of paternal smoking on adolescent behavior [20], while other studies show that friends, especially best friends, significantly influence high school students' smoking habits [21]. Furthermore, adolescents with many smoking classmates are nine times more likely to smoke [22]. Other researchers emphasize the strong correlation between adolescents' smoking habits and those of their close peers [23].

In our study, we analyze three distinct groups—current smokers, former smokers, and non-smokers—to understand smoking behavior among the adult population in rural Eastern Europe and to inform new prevention strategies. Most existing studies have focused exclusively on current smokers [24,25], with only a few discussing the initiation of smoking by former smokers or non-smokers due to their personal network connections [26]. For example, according to [27], smoking behaviors, both initiation and cessation, can spread through social networks. We propose that *assortativity*—a concept indicating that people with similar traits (e.g., smoking status) are more likely to share social ties—provides a valuable analytical framework for understanding the clustering patterns observed in smoking behavior [28,29]. Assortativity does not separate *social selection* (e.g., people who smoke prefer to interact with people who smoke), *peer socialization* or *social influence* (e.g., people adopt smoking as a result of interacting with those who smoke), and *context* (*confounding*), where external factors related to the environment or setting influence network formation.

Studies identify influential family ties (parents and siblings) and friends. Some find peer influence outweighs parental influence [30-32], while others argue parental influence is equal or greater [33,34]. These studies underscore the importance of social networks and context in shaping smoking behavior. Understanding the roles of peers and family is crucial for controlling smoking behaviors. Personal networks (individuals and their direct social contacts) constitute the immediate social context that influences people. Examining the structure (how people are connected) and composition (whom people interact with) of social networks is useful for modeling smoking behaviors. The literature highlights that a dense social network can reinforce group behavioral norms [35]. Opinion leaders increase their influence as their centrality in networks grows [35,36]. Isolated adolescents are more likely to smoke [37,38], suggesting that fewer social connections correlate with higher smoking rates. Conversely, individuals in denser social networks are more likely to share similar smoking behaviors due to shared attitudes or social control within the network.

In our study, we explored the clustering of tobacco use within personal networks in a rural Eastern European context, deploying a personal network research design [39]. A *personal network* includes a focal individual (*ego*), their direct social contacts (*alters*), and the specific patterns of connections among those alters. This methodology enables us to study how social tie arrangements affect smoking habits in rural areas, filling crucial research gaps and highlighting local dynamics. We evaluated quantitative cross-sectional data from adults over 18 in Lerești, a small rural village in Argeș County, Romania. We collected data on 76 people's personal networks (egos), social contacts (alters), and alter connections to study network impacts on middle-aged and older adults. Our hypothesis was that smoking behaviors (e.g., active smoking) are more likely when peers (social contacts) exhibit similar smoking behaviors. The research objective was to assess the extent to which assortativity predicts patterns of tobacco use in these rural settings.

# 2. Materials and methods

## 2.1. Research design and terminology

Social network analysis (SNA) is a mixed-methods framework for investigating social relations, supported by a robust theoretical foundation [40] and comprehensive research tools [41]. There are two main approaches to studying relational patterns. The first explores and tests explanations for the generative processes of networks, identifying factors that explain observed patterns [42]. The second assesses the impact of relational patterns on outcomes such as health [43,44], idea spread [45], capitalization [46], and employment opportunities [47].

Methodologically, there are three research designs for collecting network data. First, the socio-centric approach depicts relationships among individuals in the same social unit, such as pupils in a classroom [48] or colleagues in a workplace [49]. Second, the ego-centric approach explores an individual's relationships within a socio-centric network. Third, the personal network approach samples data from multiple social environments in which a respondent is embedded [39].

In SNA, relationships are referred to as social ties, with networks consisting of nodes (individuals) and their ties. In socio-centric or ego-centric designs, data are collected from each node in the network, similar to connecting points in a graph. In a personal network analysis (PNA), respondents (*egos*) provide information about themselves, their social contacts (*alters*), and the relationships among these contacts (alter-*alter* ties). Alters are sampled from various social circles in which the ego is embedded. Unlike other approaches, PNA typically produces networks based solely on information from the egos [50,51].

## 2.2. Setting

We interviewed 83 people selected from a small rural community in Romania. Due to incomplete data, we removed seven study participants and kept the rest of the 76 individuals in the analysis. We employed a PNA research design [39]. Data were collected from study participants (dubbed *egos*). Egos reported information about themselves and nominated social contacts (*alters*) based on a name generator question. They provided information about each alter's characteristics, *alter-alter* ties, and ego-*alter* ties. There are two types of information: *attribute data* (characteristics of egos and alters) and *network* (structural) *data* (how social contacts and egos are embedded in the personal network).

The size of network generators (i.e., how many people a respondent is asked to nominate) has been a research topic. Research suggests the number of elicited alters should be between 30 and 60 for better network property capture [51-53]. Working with name generators of at least 40 imposes practical problems in terms of data collection. As the number of alters increases, the respondent burden [50,54] also increases, with all its subsequent effects. Recent work shows that 25 alters is a good trade-off between a good estimation of structural properties and respondent burden [55].

The data collection was part of a fieldwork research conducted in Lerești (a small rural community), Argeș County, Romania. We gathered the data between September 13[th] and September 30[th], 2023. The study data is unique, and comes from a cohort of respondents that were included in a living lab. We plan to observe this group of study participants at multiple points in time during 2023 – 2025. The concept of a Living Lab does not have a widely recognized definition [56]. Nevertheless, The European Network of Living Labs (ENoLL), an umbrella organization for all the living labs around the world, defines it as: "*user-centered open innovation ecosystems based on a systematic user co-creation approach, integrating research and innovation processes in real-life communities and setting*" [57].

We deployed PNA because this network research design best fits our objective of assessing whether the smoking behavior of the ego correlates with the smoking behavior of their elicited social contacts. Given the rural setting, PNA allows for the collection of network data while respecting participants' privacy, as it focuses on egos' perceptions and reported interactions rather than mapping the entire community network.

## *2.3. Data Collection*

We deployed a link-tracing sampling design to recruit the study participants [58-60]. We identified the participants in a chain-referral fashion: interviewed participants recommend other interested individuals. Based on ethnographic fieldwork notes, our recruitment commenced with six diverse seeds (initial participants) to ensure a heterogeneous mix covering a broad spectrum of socio-economic indicators and age groups, from young to older individuals. It is important to note that our sample predominantly consisted of middle and older adults. This demographic skew is reflective of the population age structure in Lerești, Argeș County, Romania, where the data collection took place, rather than a methodological bias. Following the data provided by National Institute of Statistics, out of the 4124 residents in Lerești, 1965 are aged 50 or older (48%) [61]. Thus, our seeds were different in sex (four males and two females), age (*Range* = 30, Min = 34, *Max* = 64), employment sector (one retired, three employed in public sector, one employed in the private sector, and one self-employed) and education (five with BA studies and one with high-school diploma). Two seeds declined to participate as interviewees but provided recommendations for participation. Face-to-face interviews were collected between September 13 – 23, 2023, in Lerești, Romania.

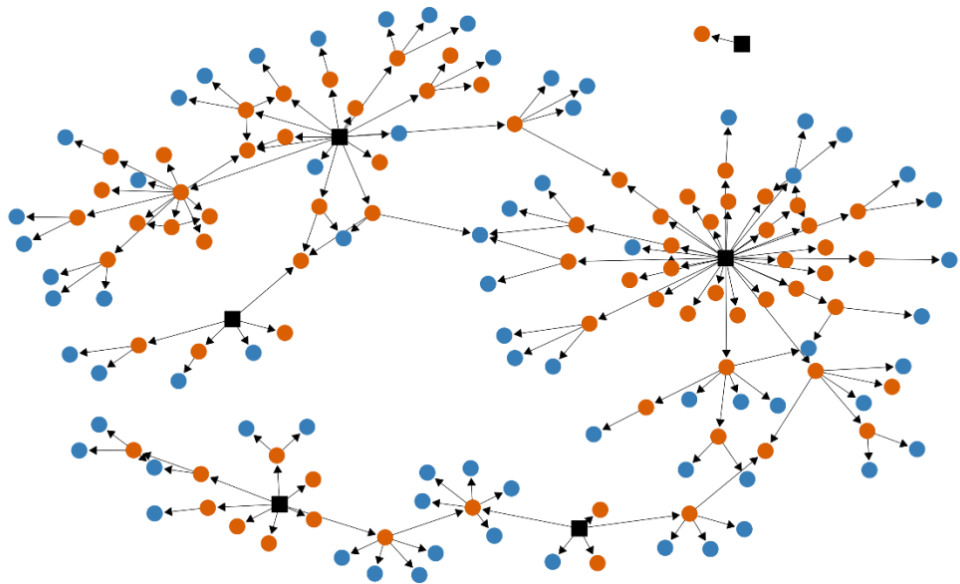

**Figure 1**. Link-tracing sampling. The black squares represent the initial individuals, referred to as 'seeds', who were invited by researchers to participate in the study. The circles depict individuals who were recommended by the study's

participants. Those marked in a tomato hue indicate individuals who participated in the study, while the ones in blue signify individuals who did not participate for various reasons, such as refusal or the need to reschedule. The directional arrows within the figure demonstrate the flow of the referral process, tracing the path from referees to their referrals.

We conducted the interviews using the Network Canvas software package [62], an advanced tool designed specifically for surveying personal networks. Respondents actively participated in the interview process, with data entry occurring in real-time within the software. This approach ensured that participants were engaged throughout the data input process, allowing them the opportunity to monitor and verify the information being recorded at any moment. The average interview lasted approximately one and a half hours. Instead of financial incentives, participants were given access to a hotline for medical second opinions and inquiries about disease prevention and healthy lifestyles. Our study followed the recommendations, guidelines, and regulations of the Romanian Sociologists Society (i.e., the professional association of Romanian sociologists) and the Declaration of Helsinki. The research protocol was approved by a named institutional/licensing committee. Specifically, the Ethics Committee of the Center for Innovation in Medicine (InoMed) reviewed and approved all these study procedures (EC-INOMED Decision No. D001/09-06-2023 and No. D001/19-01-2024). All participants gave written informed consent. The privacy rights of the study participants were observed. The authors did not have access to information that could identify participants. After each interview, information that could identify the participants was anonymized. Before conducting the interview, we provided each participant with a dossier containing informative materials about the project's objectives, how the data would be analyzed and reported, and their participation rights (e.g., the right to withdraw from the project at any time, even after the interview was completed).

## 2.4. The Questionnaire

The questionnaire was structured in several blocks. (1) *Sociodemographic variables survey*. We surveyed egos on various sociodemographic variables, including sex assigned at birth, age, marital status, education level, employment status, and smoking status (smokers, occasional smokers, former smokers, non-smokers, and never smokers). (2) *Alter generator:* Next, we incorporated an alter generator to identify individuals from the ego's close network. We requested each ego to nominate at least 25 alters (aged 18 years or older) with whom they frequently interact and feel emotionally close. We asked each ego to nominate at least 25 alters (aged 18 years or older) with whom they frequently interact and feel emotionally close. This aimed to include both close circle individuals and broader personal network ties, ensuring sufficient data for statistical modeling [63]. (3) *Name interpreter*. Egos provided sociodemographic details and smoking statuses of the nominated alters, using similar response options as for egos. (4) *Connections between alters*. Egos assessed whether their alters know and communicate with each other in the ego's absence. (5) *Intensity of ego-alter ties*. We assessed the intensity of ego-alter ties by asking about the frequency of their communications (ranging from more than once a week to less than once a year) and their level of emotional closeness (from 'not close at all' to 'very close'). We also created a binary variable for meeting frequency (at least weekly and less than weekly). Prompts included: "*How often do you typically meet with each of the people you previously mentioned?*" and "*How emotionally close do you feel to the people you previously mentioned?*"

*2.5. Attribute Data*

The study participants provided information regarding their sex (assigned at birth: male or female), date of birth, and the highest level of educational attainment. This latter was coded as a binary variable, with 0 indicating 'no university degree' and 1 indicating 'possession of a university degree'. Employment status was identified as *employed*, *unemployed*, or *retired* and subsequently coded as a binary variable, with 0 representing 'unemployed' (this category includes retired individuals) and 1 representing 'employed.' Marital status was recorded as *single*, *married*, or *in a relationship*, and later coded as a binary variable: 0 for 'single' and 1 for 'in a relationship' (this includes married individuals).

Participants were also asked about their smoking habits, with options to identify as *smokers*, *occasional smokers, former smokers, non-smokers* (including those who have smoked too infrequently to be considered regular smokers), and *never smokers*. For analysis, these responses were later grouped into three categories: *smokers* (encompassing both regular and occasional smokers), *former smokers*, and *non-smokers* (including individuals who have smoked minimally). Participants also self-reported any medical conditions (yes/no) and their participation in sports activities (yes/no). Additionally, they were asked about the occurrence of cancer or cardiovascular diseases among immediate family members (yes/no). Participants provided details about their social contacts (alters), including each alter's age, sex (assigned at birth: male or female), the highest level of education, and whether the alter adheres to a specific diet or engages in sports as a hobby (yes/no).

*2.6. Compositional Data*

For each personal network, we assessed composition by calculating the *proportion of alters* categorized as smokers, former smokers, and non-smokers, which indicates the predominance of smoking behaviors within the network. We then calculated *network density*, the ratio of actual to possible connections, to gauge the potential speed of behavior and information transmission. Lower density networks may lack interconnectedness, possibly leading to insufficient social support for healthy behaviors.

Additionally, we identified *network components*, which are sub-networks where members are interconnected but disconnected from other sub-networks. Recognizing components with higher incidences of specific smoking behaviors allows for tailored intervention strategies. Lastly, we measured *network centralization* to understand how smoking behaviors propagate through the network, focusing on key central nodes for targeted influence.

In our study, the *assortativity variable* is the variable of interest. This variable enabled us to examine the likelihood of alters' smoking behaviors being influenced by the smoking habits of their connected peers. We calculated this variable for each alter within our personal networks, employing the formula established in prior studies on assortativity related to COVID-19 vaccine uptake [64] and dietary patterns [65]. Specifically, assortativity was determined by calculating the difference between the proportion of an alter's neighbors who smoke and the overall proportion of smokers within the ego network, excluding the alter in question from this calculation. We applied the same method to compute assortativity for former smokers and non-smokers. Each category of smoking status was dichotomized (where 1 represents smokers/former smokers/non-smokers and 0 encompasses all other statuses). These computations allow us to test the hypothesis that an alter's smoking behavior is associated with the smoking patterns of their immediate social circle. For our analysis of the 76 ego networks, we excluded peers who did not have neighbors, as the assortativity measure does not apply to them.

We illustrate the calculation of the assortativity variable using smoking behavior as an example. Consider ego j's network consisting of 25 alters, including nine smokers. If we focus on a specific smoker alter *k* with four smoking neighbors, we exclude alter *k* to calculate the proportion of smokers in ego j's network, which is 8 out of 24, or 0.33. Among alter k's neighbors, the proportion of smokers is four out of four, or 1.00. The assortativity variable for alter *k* is then 0.67, representing the difference between the smoking prevalence among alter *k*'s neighbors (1.00) and the adjusted prevalence in ego *j*'s network (0.33).

## 2.7. The statistical analysis

We used a multilevel analytic approach to predict the smoking status of alters, considering the hierarchical structure where alters (level one) are nested within egos (level two). The dependencies within networks were also accounted for in our analysis. The dependent variable was the binary-coded smoking status of alters: '1' for smokers and '0' for non-smokers; '1' for former smokers and '0' for non-former smokers; '1' for non-smokers and '0' for non-non-smokers.

Actor-level variables for each ego included *sex* (coded 0 for male, 1 for female); *age* (in years, > 18); *education level* (categorized according to International Standard Classification of Education [ISCED] levels); and *employment status* (1 for employed, 2 for unemployed, 3 for retired). *Marital status* was coded as 1 for married, 2 for in a relationship, 3 for single. Smoking status was categorized as 1 for smoker, 2 for occasional smoker, 3 for former smoker, 4 for non-smoker (including minimal smokers), and 5 for never smokers.

Since the alters were not directly interviewed, we relied on the egos' perceptions as proxies to gather information about the alters. This approach involved asking the egos to provide details on the alters' *sex*, *age*, *education level*, *marital status*, and *smoking status*, using the following prompt: '*Of the persons you mentioned, please tell me which of them are* [...].

Furthermore, we centralized the *age* variable for both egos and alters to improve the numerical stability of our models, especially to mitigate the impact of outliers. Subsequently, we calculated key structural network metrics: *density*, which reflects the proportion of actual connections relative to the maximum possible connections within the network; *degree centrality*, indicating the number of direct connections a node has with others; and *the number of components*, identifying distinct interconnected groups within the network that are isolated from each other. We also performed centralization adjustments on degree and betweenness centrality metrics to account for their varying importance across networks of different sizes.

In our analysis, we examined the personal networks of 76 egos, which included a total of 1681 alters. For each ego, we assigned a binary variable to denote their *smoking status*: '1' was used to indicate a *current smoker*, while '0' represented non-smokers and other categories. This coding scheme was applied similarly to distinguish *former smokers* and *non-smokers*. An analogous binary variable was also assigned for each alter to reflect their smoking status. Our primary objective was to deploy models to predict the smoking behaviors of alters based on these binary classifications.

## 2.8. Data Exclusion

Our sample consisted of 83 respondents, characterized by varying sizes of social networks and incomplete data regarding the smoking status of the alters. For analytical rigor, we refined our sample to include only 76 individuals with personal networks comprising at least 20 alters. This selection criterion was guided by both statistical needs and the necessity to exclude isolated alters, as their presence would preclude the computation of assortativity scores, which rely on the existence of connections among network members.

## 3. Results

### 3.1. Descriptive statistics

Table 1 outlines the sociodemographic attributes of the egos. Among our 76 participants, the majority are non-smokers (either never smoked or smoked minimally) ($f= 32$; 42.1%). The distribution by sex is balanced ($f= 38$; 50% for both males and females), with a slight majority being part of the active workforce ($f= 40$; 52.6%). The vast majority are married ($f= 66$; 86.8%), with a median age of 55 years (Range = 62.0). Educationally, most have completed education beyond high school (non-tertiary level) ($f= 44$; 57.9%). Summary statistics for personal networks' structural features show that the typical personal network comprises roughly two distinct components (Median = 2.0, Range = 16.0), likely reflecting family and close friends, characterized by a low centralization score (Mean = 0.4; SD =0.1) and relatively low network density (Mean = 0.3; SD = 0.2). Table 1 also provides the sociodemographic characteristics of the 1681 alters, who are predominantly female ($f= 886$; 52.7%), either married or in a relationship ($f= 1347$; 80.1%) and have also predominantly completed education beyond high school ($f= 1001$; 59.5%). Their median age is comparable to that of the egos (Median = 53 years, Range = 77). The table includes the frequency of meetings between ego and alters, indicating that these interactions occur at least weekly ($f= 948$; 56.4%). The network's structural characteristics reveal that alters generally have a low degree of connectivity within the network (Mean = 0.05; SD = 0.03). Additionally, alters tend to play a fairly average role in linking others within the network, as indicated by the betweenness centrality score (Mean = 0.4; SD = 0.09). Regarding the assortativity variable, the former smoker category shows the highest standard deviation (SD = 0.22) compared to the smoker (SD = 0.14) and non-smoker (SD = 0.20) categories. In this sense, the network exhibits the most variability in how former smokers are connected to the egos.

| Characteristic | | Values for ego & network data (n = 76) | Values for alter data (n = 1681) |
|---|---|---|---|
| Age in years, mean (SD) | | 54.04 (16.1) | 52 (16.1) |
| Median age (Range) | | 55 (18-80) | 53 (18- 95) |
| **Smoking status, n (%)** | | | |
| | Smokers | 20 (26.3%) | 455 (27.1%) |
| | Former smokers | 24 (31.6%) | 190 (11.3%) |
| | Non-smokers | 32 (42.1%) | 1036 (61.6%) |
| **Sex, n (%)** | | | |
| | Male | 38 (50.0%) | 795 (47.3%) |
| | Female | 38 (50.0%) | 886 (52.7%) |
| **Education level, n (%)** | | | |
| | Lower education level (no university degree) | 44 (57.9%) | 1001 (59.5%) |
| | Higher education level (at least university degree) | 32 (42.1%) | 622 (37.0%) |
| | Missing | 0 (0.0%) | 58 (3.5%) |
| **Employment status, n (%)** | | | |
| | Employed | 40 (52.6%) | - |
| | Unemployed (including retired people) | 36 (47.4%) | - |
| **Marital status, n (%)** | | | |
| | Single | 10 (13.2%) | 333 (19.9%) |
| | In a relationship (married or not) | 66 (86.8%) | 1347 (80.1%) |
| Mean components (SD) | | 3.5 (3.6) | |
| Median components | | 2 | |
| Mean degree centralization (SD) | | 0.4 (0.1) | |
| Mean density (SD) | | 0.3 (0.2) | |
| Mean alter degree (SD) (mean centered and scaled) | | - | 0.05 (0.03) |
| Mean alter betweenness (SD) (mean centered and scaled) | | - | 0.04 (0.09) |
| Alters' assortativity score for smoker status | | - | 0.01 (0.14) |
| Alters' assortativity score for former smoker status | | - | -0.01 (0.22) |
| Alters' assortativity score for non-smoker status | | - | 0.02 (0.20) |

**Table 1.** Descriptive statistics for egos and alters.

Table 2 illustrates the distribution of egos by their smoking status and various demographic variables. Namely, the distribution of smokers is balanced across gender (*f*= 10; 50.0%). A significant proportion of smokers (*f*= 18; 90.0%) report being in a relationship, with an equal division across educational attainment levels, categorized as lower (less than a university degree) and higher education (at least a university Degree). Furthermore, the majority of these individuals are employed (*f*= 15; 75.0%). The mean age for smokers among the egos is the youngest across all groups, averaging 49.1 years. The table also includes the type of relationship that the ego has with each nominated alter (i.e., family member, close friend, simple friend, or acquaintance). The results indicate that, regardless of the ego's smoking status, most connections are with alters who are considered family members.

Among the egos who are former smokers, a significant majority are male ($f= 19$; 79.2%) and are currently in a relationship ($f= 21$; 87.5%). A notable proportion is unemployed ($f= 13$; 54.2%) and possesses a lower educational attainment ($f= 14$; 58.3%). When compared to current smoking egos, former smokers exhibit an older median age (Median = 58 years, Range = 58.0). Similarly, within the alter group, former smokers are predominantly males ($f= 134$; 70.5%). Most possess lower educational qualifications ($f= 122$; 65.6%) and are in a relationship ($f= 164$; 86.3%). The average age among alters who have quit smoking stands at 55.6 years. This aligns with research indicating an increased likelihood of smoking cessation with advancing age [66,67].

| Characteristic | | Smoking status | | |
|---|---|---|---|---|
| Egos | | Smokers (including occasional smokers) | Former smoker | Never-smoker (including smoked too little) |
| Age in years, mean (SD) | | 49.1 (13.8) | 54.7 (16.2) | 56.5 (17.2) |
| Median age (Range) | | 51.5 (18-68) | 58 (22-80) | 65.5 (21-76) |
| Sex, n (%) | | | | |
| | Male | 10 (50.0%) | 19 (79.2%) | 9 (28.1 %) |
| | Female | 10 (50.0%) | 5 (20.8%) | 23 (71.9%) |
| Education level, n (%) | | | | |
| | Lower education level (no university degree) | 10 (50.0%) | 14 (58.3%) | 20 (62.5%) |
| | Higher education level (at least university degree) | 10 (50.0%) | 10 (41.7%) | 12 (37.5%) |
| Employment status | | | | |
| | Unemployed | 5 (25.0%) | 13 (54.2%) | 18 (56.2%) |
| | Employed | 15 (75.0%) | 11 (45.8%) | 14 (43.8%) |
| | | | | |
| Marital status, n (%) | | | | |
| | Single | 2 (10.0%) | 3 (12.5%) | 5 (15.6%) |
| | In a relationship (including married) | 18 (90.0%) | 21 (87.5%) | 27 (84.4%) |
| Type of ego relationship with alters | | | | |
| | Family member | 216 (48.5%) | 226 (41.8%) | 394 (56.7%) |
| | Close friend | 118 (26.5%) | 107 (19.8%) | 115 (16.5%) |
| | Simple friend | 93 (20.9%) | 128 (23.7%) | 128 (18.4%) |
| | Acquaintance | 18 (4%) | 80 (14.8%) | 58 (8.3%) |

**Table 2**. Descriptive statistics based on smoking status (egos case). Note: the table shows the descriptive statistics for three categories of smoking status. They are grouped as follows: smokers (which includes current and occasional smokers), former smokers and non-smokers (respondents who did not smoke or smoked too little to be considered smokers).

The majority of our non-smoking participants are female ($f= 23$; 71.9%) and either married or in a relationship ($f= 27$; 84.4%). When it comes to educational attainment, a significant portion primarily falls within the lower education category ($f= 20$; 62.5%). Among the never-smoking egos, this group emerges as the oldest, with an average age of 56.5 years and a median age of 65.5 years, spanning an age range of 21 to 76 years. This trend may suggest that the propensity to never smoke or to smoke minimally escalates with advancing age. Similarly, never-smoking alters display a mean age of 53.5 years and a median age of 55 years, with their ages varying from 18 to 95 years. This indicates that, akin to their ego counterparts, alters who have never smoked also tend to be older, albeit with a wide age distribution.

In alignment with ego data, the data from Table 3 show a substantial majority of the smoking alters are also in a relationship ($f= 362; 79.6\%$), with a significant number having achieved a lower education level (non-university education) ($f= 265; 60.6\%$), whereas 39.4% have attained higher education (at least a bachelor's degree). Alters exhibit a slightly younger demographic compared to smoking egos, with a median age of 50 years (Range = 64.0) versus the smoking egos' median age of 51.5 years (Range = 50.0). When examining the frequency of interactions between alters and egos, it becomes evident that the majority of alters, irrespective of their smoking status, tend to meet with egos at least on a weekly basis.

| Characteristic | | Smoking status | | |
|---|---|---|---|---|
| **Alters** | | Smokers (including occasional smokers) | Former smoker | Never-smoker (including smoked too little) |
| Age in years, mean (SD) | | 46.9 (14.4) | 55.6 (13.3) | 53.5 (16.8) |
| Median age (Range) | | 50 (18-82) | 56 (21-85) | 55 (18-95) |
| **Sex, n (%)** | | | | |
| | Male | 253 (55.6%) | 134 (70.5%) | 408 (39.4%) |
| | Female | 202 (44.4%) | 56 (29.5%) | 628 (60.6%) |
| **Education level, n (%)** | | | | |
| | Lower education level (no university degree) | 265 (60.6%) | 122 (65.6%) | 614 (61.3%) |
| | Higher education level (at least university degree) | 172 (39.4%) | 64 (34.4%) | 386 (38.6%) |
| | Missing | 18 (4%) | 4 (2.1%) | 36 (3.5%) |
| **Marital status, n (%)** | | | | |
| | Single | 93 (20.4%) | 26 (13.7%) | 215 (20.8%) |
| | In a relationship (including married) | 362 (79.6%) | 164 (86.3%) | 821 (79.2%) |
| **Ego-alter meeting frequency, n (%)** | | | | |
| | Less than weekly | 200 (44.00%) | 73 (38.4%) | 460 (44.4%) |
| | At least weekly | 255 (56.00%) | 117 (61.6%) | 576 (55.6%) |

**Table 3**. Descriptive statistics based on smoking status (alter case). Note: the table shows the descriptive statistics for three categories of smoking status. They are grouped as follows: smokers (which includes current and occasional smokers), former smokers and non-smokers (respondents who did not smoke or smoked too little to be considered smokers).

### *3.2. Model estimation*

We conducted a multilevel regression analysis, creating separate models for each smoking category. The structure of the analysis comprises four distinct models: Model 0, which establishes a baseline for comparisons; Model 1, incorporating individual attributes; Model 2, examining network characteristics; and Model 3, a comprehensive model that integrates both individual attributes and network characteristics. Table 4 reports the results of fitting models that predict alters smoking behavior. The dependent variable in our analysis is the smoking status of the alter, coded as a binary variable (smoker versus all other categories). For *smokers* and *current smokers*, the findings indicate a reduced likelihood of smoking associated with being female (Model 1, Odds Ratio [OR] = 0.61; 95% Confidence Interval [CI]: 0.48, 0.78, $p < 0.001$; Model 3, OR = 0.60; CI:

0.47, 0.77, p < 0.001) and increased age (Model 1, OR = 0.68; 95% CI: 0.59, 0.78, p < 0.001; Model 3, OR = 0.68; CI: 0.59, 0.79, p < 0.001).

| | Model 0 ('intercept') | | Model 1 ('attributes') | | Model 2 ('network') | | Model 3 ('full') | |
|---|---|---|---|---|---|---|---|---|
| | OR (CI) | P | OR (CI) | P | OR (CI) | P | OR (CI) | P |
| Alter sex [ref. = male] | | | **0.61** **0.48, 0.78** | **<0.001** | | | **0.60** **0.47, 0.77** | **<0.001** |
| Alter age (mean centered, scaled) | | | **0.68** **0.59, 0.78** | **<0.001** | | | **0.68** **0.59, 0.79** | **<0.001** |
| Alter education [ref. = lower education] | | | 0.97 0.74, 1.26 | 0.793 | | | 1.00 0.77, 1.31 | 0.998 |
| Alter marital status [ref. = single] | | | 0.95 0.70, 1.29 | 0.742 | | | 0.96 0.70, 1.31 | 0.780 |
| Ego sex [ref. = male] | | | 0.86 0.61, 1.21 | 0.386 | | | 0.81 0.57, 1.15 | 0.235 |
| Ego age (mean centered, scaled) | | | 0.95 0.79, 1.14 | 0.585 | | | 0.92 0.75, 1.12 | 0.403 |
| Ego education [ref. = lower education] | | | 0.73 0.51, 1.03 | 0.075 | | | 0.72 0.50, 1.05 | 0.087 |
| Ego marital status [ref. = single] | | | 0.93 0.57, 1.51 | 0.763 | | | 0.95 0.56, 1.60 | 0.844 |
| Ego employment [ref. = unemployed] | | | 1.17 0.80, 1.70 | 0.426 | | | 1.15 0.77, 1.70 | 0.500 |
| Acquaintance smoker [ref. = other] | | | 2.61 0.85, 8.02 | 0.095 | | | 2.55 0.81, 7.98 | 0.109 |
| Simple friend smoker [ref. = other] | | | 1.52 0.85, 2.70 | 0.154 | | | 1.45 0.79, 2.65 | 0.226 |
| Close friend smoker [ref. = other] | | | **2.61** **1.55, 4.40** | **<0.001** | | | **2.47** **1.44, 4.23** | **0.001** |
| Family member smoker [ref. = other] | | | **2.64** **1.73, 4.04** | **<0.001** | | | **2.51** **1.62, 3.91** | **<0.001** |
| Ego alter meeting frequency [ref. = less than weekly] | | | | | 0.99 0.77, 1.28 | 0.946 | 1.07 0.82, 1.41 | 0.606 |
| Alter degree (mean centered, scaled) | | | | | 1.07 0.92, 1.25 | 0.359 | 1.01 0.86, 1.19 | 0.882 |
| Alter betweenness (mean centered, scaled) | | | | | 1.00 0.87, 1.16 | 0.951 | 1.02 0.88, 1.18 | 0.811 |
| Network components | | | | | 1.00 0.92, 1.09 | 0.986 | 1.04 0.96, 1.12 | 0.349 |
| Network degree centralization | | | | | 0.91 0.19, 4.33 | 0.908 | 1.00 0.24, 4.11 | 0.999 |
| Network density | | | | | 3.28 0.79, 13.60 | 0.101 | 1.68 0.47, 6.03 | 0.426 |
| Smoking assortativity (smokers) | | | | | **3.44** **1.85, 6.37** | **<0.001** | **3.01** **1.61, 5.62** | **0.001** |
| Smoking assortativity (former smokers) | | | | | 1.83 0.73, 4.56 | 0.198 | 1.57 0.63, 3.91 | 0.334 |
| Intercept | **0.34** **0.28, 0.41** | **<0.001** | **0.42** **0.24, 0.75** | **0.003** | **0.22** **0.07, 0.70** | **0.010** | **0.30** **0.10, 0.89** | **0.030** |
| Num.obs | 1622 | | 1622 | | 1622 | | 1622 | |
| Num.groups: ego_id | 76 | | 76 | | 76 | | 76 | |
| ICC | 0.13 | | 0.06 | | 0.12 | | 0.07 | |
| AIC | 1836.164 | | 1775.478 | | 1831.758 | | 1777.152 | |
| BIC | 1846.947 | | 1856.349 | | 1885.672 | | 1901.155 | |
| Log Likelihood | -916.082 | | -872.739 | | -905.879 | | -865.576 | |
| Deviance | 1832.164 | | 1745.478 | | 1811.758 | | 1731.152 | |
| Marginal R2 / Conditional R2 | 0.000 / 0.127 | | 0.117 / 0.173 | | 0.026 / 0.143 | | 0.131 / 0.188 | |

**Table 4**. Multilevel logistic regression models predicting alters' being smokers.

The presence of a family member who smokes (Model 1, Odds Ratio [OR] = 2.64; 95% Confidence Interval [CI]: 1.73, 4.04, $p < 0.001$; Model 3, OR = 2.51; CI: 1.62, 3.91, $p < 0.001$) or a close friend who smokes (Model 2, OR = 2.61; 95% CI: 1.55, 4.40, $p < 0.001$; Model 3, OR = 2.47; CI: 1.44, 4.23, $p < 0.05$) significantly increases the likelihood of an individual being a smoker. Among network-related variables, the *assortativity score* stands out as the sole significant predictor, demonstrating a consistent positive relationship across the models (Model 2, OR = 3.44; 95% CI: 1.85, 6.37, $p < 0.001$; Model 3, OR = 3.01; CI: 1.61, 5.62, $p < 0.001$). This finding suggests that alters sharing similar smoking habits tend to form clusters within the network, indicating that smokers are more inclined to have connections with other smokers, particularly among family members and close friends, with family ties showing a more pronounced impact.

In the analysis of former smokers (Table 5), female sex significantly reduces the likelihood of being a former smoker (Model 1, Odds Ratio [OR] = 0.33; Confidence Interval [CI]: 0.22, 0.48, $p < 0.001$; Model 3, OR = 0.31; CI: 0.21, 0.45, $p < 0.001$). Conversely, advancing age substantially increases the probability of being a former smoker (Model 1, OR = 1.76; CI: 1.41, 2.19, $p < 0.001$; Model 3, OR = 1.75; CI: 1.40, 2.19, $p < 0.001$). Among the ego characteristics, only age shows a significant effect (Model 1, OR = 0.65; CI: 0.44, 0.96, $p < 0.05$; Model 3, OR = 0.56; CI: 0.38, 0.82, $p < 0.05$). This suggests that unlike in the case of current smokers where both family and close friends' smoking statuses were influential, for former smokers, only the presence of close friends who are former smokers significantly influences one's own former smoker status (Model 3, OR = 2.82; CI: 1.08, 7.44, $p < 0.05$).

The *assortativity score* for former smokers serves as a potent predictor of clustering behavior, as evidenced in Model 2 (Odds Ratio [OR] = 6.34; Confidence Interval [CI]: 2.35, 17.09, $p < 0.001$) and Model 3 (OR = 4.59; CI: 1.60, 13.13, $p < 0.05$), indicating a tendency for former smokers to form clusters within social networks based on similar smoking statuses. Additionally, our analysis extends to the assortativity score for current smokers, which also demonstrates significance (Model 2, OR = 2.67; CI: 1.04, 6.85, $p < 0.005$; Model 3, OR = 4.00; CI: 1.50, 10.67, $p < 0.05$). This finding highlights the influence of current smokers within networks of former smokers, suggesting that in networks characterized by prevalent smoking, there could be an intensified peer influence or social pressure that potentially encourages former smokers to resume smoking.

The situation of former smokers, particularly when compared to current smokers, underscores the significance of network metrics. In this vein, the finding regarding network degree centralization is noteworthy: it indicates that in networks characterized by a few individuals having significantly more connections than others, there exists a higher likelihood of alters being former smokers (Model 3, Odds Ratio [OR] = 24.19; Confidence Interval [CI]: 1.41, 415.64, $p < 0.05$). This outcome suggests the pivotal role of certain network members who may exert influence on the smoking behaviors of former smokers, potentially aiding in their cessation efforts or, conversely, contributing to a higher risk of relapse. The broad confidence interval for this estimate, however, warrants a cautious interpretation of these findings. Furthermore, the analysis highlights the relevance of the number of components (Model 2, OR = 1.17; CI: 1.00, 1.36, $p < 0.05$; Model 3, OR = 1.23; CI: 1.06, 1.42, $p < 0.05$), indicating that engagement with diverse social groups can have an impact on smoking behavior. Also, the finding regarding network degree centralization is noteworthy: it indicates that in networks characterized by a few individuals having significantly more connections than others, there exists a higher likelihood of alters being former smokers (Model 3, Odds Ratio [OR] = 24.19; Confidence Interval [CI]: 1.41, 415.64, $p < 0.05$).

|  | Model 0 ('intercept') | | Model 1 ('attributes') | | Model 2 ('network') | | Model 3 ('full') | |
| --- | --- | --- | --- | --- | --- | --- | --- | --- |
|  | OR (CI) | P | OR (CI) | P | OR (CI) | P | OR (CI) | P |
| Alter sex [ref. = male] |  |  | **0.33** **0.22, 0.48** | **<0.001** |  |  | **0.32** **0.22, 0.47** | **<0.001** |
| Alter age (mean centered, scaled) |  |  | **1.76** **1.41, 2.19** | **<0.001** |  |  | **1.75** **1.40, 2.19** | **<0.001** |
| Alter education [ref. = lower education] |  |  | 0.89 0.60, 1.32 | 0.565 |  |  | 0.92 0.62, 1.38 | 0.694 |
| Alter marital status [ref. = single] |  |  | 1.49 0.91, 2.44 | 0.111 |  |  | 1.43 0.87, 2.35 | 0.161 |
| Ego sex [ref. = male] |  |  | 1.05 0.48, 2.28 | 0.899 |  |  | 1.04 0.50, 2.16 | 0.910 |
| Ego age (mean centered, scaled) |  |  | **0.65** **0.44, 0.96** | **0.030** |  |  | **0.56** **0.38, 0.82** | **0.003** |
| Ego education [ref. = lower education] |  |  | 0.88 0.43, 1.80 | 0.732 |  |  | 0.94 0.47, 1.87 | 0.857 |
| Ego marital status [ref. = single] |  |  | 0.82 0.31, 2.20 | 0.699 |  |  | 0.80 0.30, 2.13 | 0.653 |
| Ego employment [ref. = unemployed] |  |  | 1.82 0.83, 3.98 | 0.132 |  |  | 1.76 0.83, 3.71 | 0.138 |
| Acquaintance former smoker [ref. = other] |  |  | 0.68 0.22, 2.09 | 0.497 |  |  | 1.03 0.34, 3.12 | 0.964 |
| Simple friend former smoker [ref. = other] |  |  | 0.66 0.23, 1.85 | 0.426 |  |  | 0.83 0.30, 2.33 | 0.724 |
| Close friend former smoker [ref. = other] |  |  | 2.27 0.86, 5.95 | 0.097 |  |  | **2.84** **1.08, 7.44** | **0.034** |
| Family member former smoker [ref. = other] |  |  | 1.55 0.66, 3.62 | 0.314 |  |  | 1.93 0.85, 4.39 | 0.118 |
| Ego alter meeting frequency [ref. = less than weekly] |  |  |  |  | 1.31 0.90, 1.92 | 0.156 | 1.12 0.75, 1.68 | 0.566 |
| Alter degree (mean centered, scaled) |  |  |  |  | 1.08 0.88, 1.33 | 0.477 | 1.02 0.82, 1.28 | 0.841 |
| Alter betweenness (mean centered, scaled) |  |  |  |  | 1.02 0.84, 1.23 | 0.839 | 1.07 0.87, 1.30 | 0.526 |
| Network components |  |  |  |  | **1.17** **1.00, 1.36** | **0.044** | **1.24** **1.07, 1.43** | **0.005** |
| Network degree centralization |  |  |  |  | 10.37 0.65, 166.24 | 0.098 | **24.19** **1.41, 415.64** | **0.028** |
| Network density |  |  |  |  | 11.43 0.86, 152.14 | 0.065 | 10.99 0.80, 150.83 | 0.073 |
| Smoking assortativity (smokers) |  |  |  |  | **2.67** **1.04, 6.85** | **0.041** | **4.00** **1.50, 10.67** | **0.006** |
| Smoking assortativity (former smokers) |  |  |  |  | **6.34** **2.35, 17.09** | **<0.001** | **4.59** **1.60, 13.13** | **0.005** |
| Intercept | 0.08 0.05, 0.12 | <0.001 | 0.07 0.02, 0.26 | <0.001 | 0.01 0.00, 0.06 | <0.001 | 0.00 0.00, 0.05 | <0.001 |
| Num.obs | 1622 |  | 1622 |  | 1622 |  | 1622 |  |
| Num.groups: ego_id | 76 |  | 76 |  | 76 |  | 76 |  |
| ICC | 0.32 |  | 0.29 |  | 0.30 |  | 0.24 |  |
| AIC | 1074.405 |  | 1016.314 |  | 1066.666 |  | 1007.600 |  |
| BIC | 1085.188 |  | 1097.185 |  | 1120.58 |  | 1131.602 |  |
| Log Likelihood | -535.203 |  | -493.157 |  | -523.333 |  | -480.800 |  |
| Deviance (df.resid) | 1070.405 |  | 986.314 |  | 1046.666 |  | 961.600 |  |
| Marginal R2 / Conditional R2 | 0.000 / 0.324 |  | 0.155 / 0.404 |  | 0.060 / 0.338 |  | 0.231 / 0.417 |  |

**Table 5**. Multilevel logistic regression models explaining alters' being former smokers.

Table 6 showcases the findings pertaining to non-smokers. Sex plays a significant role, with being female markedly elevating the likelihood of being a non-smoker (Model 1, Odds Ratio [OR] =

2.49; 95% Confidence Interval [CI]: 1.97, 3.16, p < 0.001; Model 3, OR = 2.56; CI: 2.01, 3.25, p < 0.001). Alter age, alter and ego education or alter and ego marital status are not statistically significant. This means that in the case of non-smokers, the sociodemographic profile has a lower importance in contrast to network characteristics. Among non-smokers, the employment status of the ego is notably significant, indicating that alters associated with employed egos are less likely to smoke compared to those connected with unemployed egos (Model 1, OR = 0.63; 95% CI: 0.40, 0.99, p < 0.05; Model 3, OR = 0.63; CI: 0.40, 0.99, p < 0.05). Additionally, ego age is significant in Model 3, suggesting that older egos may exert a greater influence on the non-smoking behavior of their alters (OR = 1.38; CI: 1.10, 1.73, p = 0.005). In other words, as ego age increases, the likelihood of influencing their social circle towards non-smoking behaviors appears to grow, underscoring the role of older individuals in promoting healthier behaviors within their networks.

The presence of a non-smoking family member significantly increases the odds of an individual being a non-smoker (Model 2, Odds Ratio [OR] = 1.83; 95% Confidence Interval [CI]: 1.15, 2.93, p < 0.05; Model 3, OR = 1.64; CI: 1.04, 2.61, p < 0.05), underscoring the role of immediate family in influencing smoking behaviors, akin to the patterns observed in smoker models. The *assortativity score* for non-smokers registers as highly significant (Model 2, OR = 5.61; CI: 2.46, 12.80, p < 0.001; Model 3, OR = 4.91; CI: 2.10, 11.47, p < 0.001), signifying a pronounced tendency for non-smokers to form clusters within their social networks. This pronounced assortativity among non-smokers accentuates the impact of social networks, indicating that non-smokers are likely influenced by the smoking habits of their peers. Furthermore, the significance of network components suggests that as the number of distinct network components increases, the likelihood of being a non-smoker slightly decreases (Model 3, OR = 0.89; CI: 0.81, 0.97, p < 0.05)

In addition to the multilevel logistic regression models, a standard logistic General Linear Model (GLM) was executed. The outcomes for all three categories of smoking behavior (smokers, former smokers, and non-smokers) are documented in Table S36, S44, S52 in the Supplementary Material. The comparison of these regression models reveals consistent results concerning the alters' sex (e.g., for smokers: OR = 0.60; CI: 0.47, 0.77, p < 0.001), age (e.g., for smokers: OR = 0.70; CI: 0.61, 0.80, p < 0.001), and assortativity score (e.g., for smokers: OR = 0.21; CI: 0.07, 0.65, p < 0.05) across specific smoking statuses. To accommodate for the absence of random ego intercepts, this model also factored in the proportion of smokers/former smokers or non-smokers excluding the alter of interest, revealing a positive and significant effect across all three smoking categories.

|  | Model 0 ('intercept') | | Model 1 ('attributes') | | Model 2 ('network') | | Model 3 ('full') | |
| --- | --- | --- | --- | --- | --- | --- | --- | --- |
|  | OR (CI) | P | OR (CI) | P | OR (CI) | P | OR (CI) | P |
| Alter sex [ref. = male] |  |  | **2.49** **1.97,3.16** | **<0.001** |  |  | **2.56** **2.01,3.25** | **<0.001** |
| Alter age (mean centered, scaled) |  |  | 1.14 1.00,1.30 | 0.052 |  |  | 1.13 0.99,1.29 | 0.069 |
| Alter education [ref. = lower education] |  |  | 1.11 0.86,1.43 | 0.417 |  |  | 1.05 0.81,1.35 | 0.731 |
| Alter marital status [ref. = single] |  |  | 0.91 0.68,1.22 | 0.533 |  |  | 0.93 0.69,1.25 | 0.613 |
| Ego sex [ref. = male] |  |  | 0.98 0.63,1.52 | 0.919 |  |  | 1.09 0.71,1.68 | 0.693 |
| Ego age (mean centered, scaled) |  |  | 1.25 1.00,1.57 | 0.053 |  |  | **1.38** **1.10,1.73** | **0.005** |
| Ego education [ref. = lower education] |  |  | 1.32 0.86,2.02 | 0.208 |  |  | 1.30 0.85,1.99 | 0.228 |
| Ego marital status [ref. = single] |  |  | 1.02 0.56,1.83 | 0.954 |  |  | 1.06 0.58,1.93 | 0.854 |
| Ego employment [ref. = unemployed] |  |  | **0.63** **0.40,0.99** | **0.043** |  |  | **0.63** **0.40,0.99** | **0.046** |
| Acquaintance non-smoker [ref. = other] |  |  | 1.51 0.69,3.31 | 0.304 |  |  | 1.35 0.61,3.00 | 0.462 |
| Simple friend non-smoker [ref. = other] |  |  | 1.08 0.59,1.98 | 0.793 |  |  | 1.06 0.58,1.94 | 0.840 |
| Close friend non-smoker [ref. = other] |  |  | 1.28 0.69,2.34 | 0.433 |  |  | 1.06 0.58,1.95 | 0.847 |
| Family member non-smoker [ref. = other] |  |  | **1.83** **1.15,2.93** | **0.011** |  |  | **1.64** **1.04,2.61** | **0.035** |
| Ego alter meeting frequency [ref. = less than weekly] |  |  |  |  | 0.90 0.71,1.15 | 0.389 | 0.94 0.73,1.22 | 0.665 |
| Alter degree (mean centered, scaled) |  |  |  |  | 0.92 0.80,1.06 | 0.256 | 0.90 0.77,1.05 | 0.167 |
| Alter betweenness (mean centered, scaled) |  |  |  |  | 0.98 0.86,1.12 | 0.814 | 0.98 0.86,1.13 | 0.804 |
| Network components |  |  |  |  | 0.95 0.86,1.04 | 0.289 | **0.89** **0.81,0.97** | **0.009** |
| Network degree centralization |  |  |  |  | 0.55 0.10,3.12 | 0.501 | 0.27 0.06,1.34 | 0.109 |
| Network density |  |  |  |  | **0.18** **0.04,0.90** | **0.037** | 0.23 0.05,1.02 | 0.053 |
| Smoking assortativity (smokers) |  |  |  |  | 1.29 0.53,3.12 | 0.579 | 1.10 0.44,2.76 | 0.834 |
| Smoking assortativity (non-smokers) |  |  |  |  | **5.61** **2.46,12.80** | **<0.001** | **4.91** **2.10,11.47** | **<0.001** |
| Intercept | 1.69 1.36,2.11 | **<0.001** | 1.07 0.53,2.14 | 0.855 | 4.89 1.36,17.53 | **0.015** | 4.29 1.24,14.89 | **0.022** |
| Num.obs | 1622 | | 1622 | | 1622 | | 1622 | |
| Num.groups: ego_id | 76 | | 76 | | 76 | | 76 | |
| ICC | 0.18 | | 0.13 | | 0.17 | | 0.11 | |
| AIC | 2056.081 | | 1987.822 | | 2029.770 | | 1957.717 | |
| BIC | 2066.864 | | 2068.693 | | 2083.684 | | 2081.72 | |
| Log Likelihood | -1026.041 | | -978.911 | | -1004.885 | | -955.859 | |
| Deviance (df.resid) | 2052.081 | | 1957.822 | | 2009.770 | | 1911.717 | |
| Marginal R2 / Conditional R2 | 0.000 / 0.176 | | 0.121 / 0.237 | | 0.046 / 0.204 | | 0.168 / 0.262 | |

**Table 6**. Multilevel logistic regression models explaining alters' being non-smokers.

# 4. Discussion

Our study analyzed the impact of family members, friends, and acquaintances with different smoking habits on the smoking habits of people located in a rural community using a PNA design. Deploying multilevel logistic regression models, we aimed to predict the smoking status of alters using various variables, such as alters' sex, education, age, relationship status (being single or in a relationship), betweenness centrality, meeting frequency between ego and alter (at least weekly or less than weekly), together with the assortativity in smoking behavior for level one. For level two, we used egos' sex, education, age, employment status (employed or unemployed), type of ego (family member, close friends, and acquaintances), and the smoking status of the ego. Network measurements included alter degree, alter betweenness, number of components, degree centralization, and density.

The primary finding of our research emphasizes the influence that social networks—consisting of friends and family members with diverse smoking patterns—have on the smoking behaviors of individuals. Specifically, the presence of smokers within one's network markedly increases the likelihood of an individual being a smoker. Conversely, having family members or close friends who are non-smokers or former smokers significantly boosts the odds of an individual being a non-smoker or a former smoker, respectively. Our analysis indicates that both close friends and family members play significant roles, yet their impact can differ depending on the smoking behavior in question—be it current smoking status, former smoking status, or the likelihood of being a non-smoker. The strong influence of family members suggests that smoking habits circulate more within the family instead of close friends' group.

Our current results are related to findings from other research, highlighting that individuals with family members who smoke are more likely to adopt similar smoking habits [68]. While multiple studies reported that the impact of close friends is larger than the one of the family [31,69], our results show similar trends but in the case of former smokers. For current smokers, our results highlight the higher importance of family. These contrasting results should be carefully considered as most studies related to the impact of friendship and family in smoking behavior are (a) focused mainly on adolescents and (b) not discriminating between profiles of smoking habits (smokers, former smokers or non-smokers).

For former smokers, the analysis reveals that having close friends who are former smokers significantly increases the likelihood of an individual also being a former smoker. The effect of family in this case is not statistically significant. In contrast, the influence of family members appears to be more pronounced in the models for non-smokers. While having a family member who is a non-smoker significantly increases the likelihood of an individual being a non-smoker, the effect size is smaller compared to the models for current smokers.

Secondary results relate to alters' sociodemographic characteristics. Our analysis revealed sex and age as critical determinants of smoking behavior, with females less likely to smoke or be former smokers and older individuals more inclined towards being former smokers. These results are in line with studies suggesting that the prevalence of smoking is higher in men than females [70]. Furthermore, our results are similar to the ones found in [71] and [72] where older adults who had ever smoked cigarettes had decided to quit.

Another important result relates to the effect of assortative mixing. Across all models, assortativity scores emerged as a very important predictor, highlighting the tendency of individuals to associate with peers sharing similar smoking behaviors. In other words, alters with a higher proportion of direct network neighbors that are smokers/former smokers or non-smokers, compared to the overall network proportion, had a higher chance of being classified as smokers/former smokers or non-smokers. This clustering effect within social networks—where smokers gravitate towards smokers, and non-smokers form connections with non-smokers—highlights the reinforcement of smoking habits through social connections.

In light of these results, we recommend that policies address both individual social characteristics and the social interactions shaping these behaviors. The study's examination of various smoking patterns highlights the need for policies tailored to specific smoker categories. This is particularly important because the assortativity score for former smokers is significantly related to current smokers, suggesting that the presence of current smokers in a social network can hinder cessation efforts. Therefore, network interventions should be designed to strengthen connections between non-smokers and former smokers, or between non-smokers and current smokers, creating a supportive environment that encourages quitting or maintaining cessation.

Our study had several limitations. The first one is related to the design of personal network analysis itself. Given that the data relies on the ego's own perception of alters, the reliance on ego-reported data about alters' attributes and relationships may be prone to potential inaccuracies such as the *false consensus effect* [73,74], or due to data recall [75] or social desirability biases [76,77]. In this sense, we limited the number of elicited alters to 25, trying to help the ego identify them based on how frequently they communicate and how emotionally close they are.

Given the fact that our study is based on cross-sectional data, we cannot distinguish between social selection and social influence within the context of smoking behaviors. Specifically, we cannot determine whether being part of a network with higher rates of smoking leads individuals to adopt smoking behaviors (social influence), or if individuals who already have a preference for smoking are more likely to associate with others who share similar smoking habits (social selection). Collecting longitudinal data may differentiate between these effects.

Despite these limitations, this study's unique contribution is to examine the role of social networks in influencing smoking behaviors among adults in a rural setting using a PNA approach. This study differentiates itself by focusing not only on current smokers but also on former smokers and non-smokers, providing a comprehensive view of smoking behaviors within personal networks. This approach offers a novel perspective on addressing smoking in rural communities, where social ties, along with sociodemographic characteristics may have a particularly pronounced impact on health behaviors.

Given that adults who live in rural areas are mostly overlooked [26], further research should be expanded to other similar areas in an effort to increase the understanding of smoking behavior in these areas. On the other hand, it is important that future research investigate whether the findings of the current study may be applied to health-related behaviors that are not only tied to smoking.

The supplementary materials, including the code and other relevant information, as well as the data, are openly available [78].

## Acknowledgements

The authors kindly thank Dr. Bianca Cucoş, Florin Găină, Bogdan-Adrian Vidraşcu, Simona-Elena Puncioiu, and Isabela Tincă, for their valuable contributions to this study.

# References

1. Samet JM. 2013 Tobacco smoking. The leading cause of preventable disease worldwide. *Thorac Surg Clin.* **23**, 103-112. (doi:10.1016/j.thorsurg.2013.01.009)

2. European Commission. 2023 Overview [Internet]. [cited 2024 Feb 04]. Available from: https://health.ec.europa.eu/tobacco/overview_en

3. Publications Office of the European Union, 2015. Ex post evaluation, ex-smokers campaign [Internet]. [cited 2024 May 02]. Available from: https://op.europa.eu/en/publication-detail/-/publication/b656dd28-cefa-11e5-a4b5-01aa75ed71a1

4. World Health Organization. 2021 WHO global report on trends in prevalence of tobacco use 2000–2030 [Internet]. [cited 2024 July 20]. Available from: https://www.who.int/publications/i/item/9789240088283

5. Banks E, Joshy G, Korda RJ, Stavreski B, Soga K, Egger S, et al. 2019 Tobacco smoking and risk of 36 cardiovascular disease subtypes: Fatal and non-fatal outcomes in a large prospective Australian study. *BMC Med*. **17**, 1–18. (doi:10.1186/s12916-019-1351-4)

6. Campagna D, Alamo A, Di Pino A, Russo C, Calogero AE, Purrello F, et al. 2019 Smoking and diabetes: dangerous liaisons and confusing relationships. *Diabetol Metab Syndr*. **11**, 1-12. (doi:10.1186/s13098-019-0482-2)

7. Raju P, George R, Ve Ramesh S, Arvind H, Baskaran M, Vijaya L. 2006 Influence of tobacco use on cataract development. *Br. J. Ophthalmol.* **90**, 1374–1377. (doi:10.1136/bjo.2006.097295)

8. Li LF, Chan RL, Lu L, Shen J, Zhang L, Wu WK, et al. 2014 Cigarette smoking and gastrointestinal diseases: the causal relationship and underlying molecular mechanisms (review). *Int. J. Mol. Med*. **34**, 372–380. (doi:10.3892/ijmm.2014.1786)

9. Burns DM. 2003 Tobacco-related diseases. *Semin. Oncol. Nurs.* **19**, 244–249. (doi:10.1053/j.soncn.2003.08.001)

10. Schuller HM. 2009 Is cancer triggered by altered signalling of nicotinic acetylcholine receptors? *Nat. Rev. Cancer.* **9**, 195–205. (doi:10.1038/nrc2590)

11. Gallus S, Lugo A, Liu X, Behrakis P, Boffi R, Bosetti C, et al. 2021 Who smokes in Europe? Data from 12 European countries in the TackSHS survey (2017–2018). *J. Epidemiol.* **31**, 17. (doi:10.2188/jea.je20190344)

12. Eurostat. 2019a 18.4% of EU population smoked daily in 2019 [Internet]. [cited 2024 April 03]. Available from: https://ec.europa.eu/eurostat/web/products-eurostat-news/-/edn-20211112-1

13. Eurostat. 2019b Daily smokers of cigarettes by sex, age and educational attainment level [Internet]. [cited 2024 April 08]. Available from: https://ec.europa.eu/eurostat/databrowser/view/hlth_ehis_sk3e/default/table?lang=en

14. World Health Organization. 2019 European tobacco use. Trends Report 2019 [Internet]. [cited 2024 April 05]. Available from: https://iris.who.int/handle/10665/346817

15. Buettner-Schmidt K, Miller DR, Maack B. 2019 Disparities in rural tobacco use, smoke-free policies, and tobacco taxes. *West. J. Nurs. Res*. **41**, 1184–1202. (doi:10.1177/0193945919828061)

16. Mitchell SA, Kneipp SM, Giscombe CW. 2016 Social factors related to smoking among rural, low-income women: findings from a systematic review. *Public Health Nurs*. **33**, 214–223. (doi:10.1111/phn.12233)

17. Doescher MP, Jackson JE, Jerant A, Hart LG. 2006 Prevalence and trends in smoking: a national rural study. *J. Rural. Health*. **22**, 112–118. (doi:10.1111/j.1748-0361.2006.00018.x)

18. Institutul Național de Statistică. 2023 16 noiembrie 2023 – Ziua națională fără tutun [Internet]. [cited 2024 June 06]. Available from: https://insp.gov.ro/2023/11/16/16-noiembrie-2023-ziua-nationala-fara-tutun/

19. World Health Organization. 2018 Global Adult Tobacco Survey Fact Sheet: Romania 2018 [Internet]. [cited 2024 March 15]. Available from: https://cdn.who.int/media/docs/default-source/ncds/ncd-surveillance/data-reporting/romania/gats-romania-2018-factsheet.pdf?sfvrsn=2206beb4_1&download=true

20. Iordanescu E, Iordanescu C, Draghici A. 2015 Like parents, like teenagers: a Romanian youth smoking overview. *Procedia Soc. Behav Sci*. **203**, 361–366. (doi:10.1016/j.sbspro.2015.08.308)

21. Lotrean LM, Ionut C, Mesters I, de Vries H. 2009 Factors associated with smoking among Romanian senior high school students. *Rev. Res. Soc. Interv*. **25**, 83–100.

22. Albert-Lőrincz E, Paulik E, Szabo B, Foley K, Gasparik AI. 2019 Adolescent smoking and the social capital of local communities in three counties in Romania. *Gac. Sanit.* **33**, 547–553. (doi:10.1016/j.gaceta.2018.05.009)

23. Lakon CM, Valente TW. 2012 Social integration in friendship networks: the synergy of network structure and peer influence in relation to cigarette smoking among high risk adolescents. *Soc Sci Med*. **74**, 1407–1417. (doi:10.1016/j.socscimed.2012.01.011)

24. Montes F, Blanco M, Useche AF, Sanchez-Franco S, Caro C, Tong L, et al. 2023 Exploring the mechanistic pathways of how social network influences social norms in adolescent smoking prevention interventions. *Sci. Rep.* **13**, 3017. (doi:10.1038/s41598-023-28161-7)

25. Burgess-Hull AJ, Roberts LJ, Piper ME, Baker TB. 2018 The social networks of smokers attempting to quit: An empirically derived and validated classification. *Psychol. Addict. Behav*. **32**(1):64–75. (doi:10.1037/adb0000336)

26. Yun EH, Kang YH, Lim MK, Oh JK, Son JM. 2010 The role of social support and social networks in smoking behavior among middle and older aged people in rural areas of South Korea: A cross-sectional study. *BMC Public Health*, **10**, 1-8. (doi: 10.1186/1471-2458-10-78)

27. Christakis NA, Fowler, JH. 2008 The collective dynamics of smoking in a large social network. *N. Engl. J. Med.* **358**, 2249–2258. (doi:10.1056/nejmsa0706154)

28. Newman MEJ. 2002 Assortative mixing in networks. *Phys. Rev. Lett*. **89**, 1–5. (doi:10.1103/physrevlett.89.208701)

29. Noldus R, Van Mieghem P. 2014 Assortativity in complex networks. *J. Complex Netw.* **3**, 507–542. (doi:10.1093/comnet/cnv005)

30. Allen M, Donohue WA, Griffin A, Ryan D, Turner MMM. 2003 Comparing the influence of parents and peers on the choice to use drugs: A meta-analytic summary of the literature. *Crim. Justice. Behav* **30**, 163–186. (doi:10.1177/0093854802251002)

31. Saari AJ, Kentala J, Mattila KJ. 2014 The smoking habit of a close friend or family member - How deep is the impact? A cross-sectional study. *BMJ Open* **4**, 1–6. (doi:10.1136/bmjopen-2013-003218)

32. Scragg R, Laugesen M. 2007 Influence of smoking by family and best friend on adolescent tobacco smoking: Results from the 2002 New Zealand national survey of year 10 students. *Aust. N. Z. J. Public Health*. **31**, 217–223. (doi:10.1111/j.1467-842X.2007.00051.x)

33. Bauman KE, Carver K, Gleiter K. 2001 Trends in parent and friend influence during adolescence: The case of adolescent cigarette smoking. *Addict. Behav.* **26**, 349–361. (doi:10.1016/S0306-4603(00)00110-6)

34. McGee CE, Trigwell J, Fairclough SJ, Murphy RC, Porcellato L, Ussher M, et al. 2015 Influence of family and friend smoking on intentions to smoke and smoking-related attitudes and refusal self-efficacy among 9-10 year old children from deprived neighbourhoods: a cross-sectional study. *BMC Public Health* **15**, 15–21. (doi.org/10.1186/s12889-015-1513-z)

35. Valente TW, Gallaher P, Mouttapa M. 2004 Using social networks to understand and prevent substance use: a transdisciplinary perspective. *Subst. Use. Misuse*. **39**, 1685–1712. (doi:10.1081/lsum-200033210)

36. Takagi D, Yokouchi N, Hashimoto H. 2020 Smoking behavior prevalence in one's personal social network and peer's popularity: a population-based study of middle-aged adults in Japan. *Soc. Sci. Med.* **260**, 113207. (doi:10.1016/j.socscimed.2020.113207)

37. Ennett ST, Faris R, Hipp J, Foshee VA, Bauman KE, Hussong A, et al. 2008 Peer smoking, other peer attributes, and adolescent cigarette smoking: a social network analysis. *Prev. Sci.* **9**, 88–98. (doi:10.1007/s11121-008-0087-8)

38. Seo DC, Huang Y. 2012 Systematic review of social network analysis in adolescent cigarette smoking behavior. *J. Sch. Health*. **82**, 21–27. (doi:10.1111/j.1746-1561.2011.00663.x)

39. McCarty C, Lubbers MJ, Vacca R, Molina JL. 2019 Conducting Personal Network Research: A Practical Guide. Guilford Publications, New York

40. Borgatti SP, Mehra A, Brass DJ, Labianca G. 2009 Network analysis in social sciences. *Science*, **323**(5916), 892-895. (doi:10.1126/science.1165821)

41. Wasserman S, Faust K. 1994 Social network analysis: methods and analysis. Cambridge, UK: Cambridge University Press.

42. Lusher D, Koskinen J, Robbins G. 2012 Exponential random graph models for social networks: theory, methods, and applications. Cambridge, UK: Cambridge University Press. (doi:10.1017/cbo9780511894701)

43. Burgette JM, Rankine J, Culyba AJ, Chu K-H, Carley KM. 2021 Best practices for modeling egocentric social network data and health outcomes. *HERD*. **14**, 18–34. (doi:10.1177/19375867211013772)

44. Grewal E, Godley J, Wheeler J, Tang KL. 2024 Use of social network analysis in health research: a scoping review protocol. *BMJ Open.* **14**, 1–7. (doi:10.1136/bmjopen-2023-078872)

45. Collar A. 2022 Networks and the spread of ideas in the past. Strong ties, innovation and knowledge exchange. Routledge, Taylor & Francis, New York. (doi:10.4324/9780429429217)

46. Borgatti SP, Halgin DS. 2011 On network theory. *Org. Sci.* **22**, 1168–1181. (doi:10.1287/orsc.1100.0641)

47. Granovetter MS. 1973 The strength of weak ties. *Am. J. Sociol*. **78**, 1360–1380. (doi:10.1086/225469)

48. Grunspan DZ, Wiggins BL, Goodreau SM. 2014 Understanding classrooms through social network analysis: a primer for social network analysis in education research. *CBE Life Sci. Educ.* **13**, 167–178. (doi:10.1187/cbe.13-08-0162)

49. Brass DJ, Galaskiewicz J, Greve HR, Tsai W, 2004 Taking stock of networks and organizations: A multilevel perspective. *Acad. Manag. J.* **47**, 795–817. (doi:10.2307/20159624)

50. Marsden PV. 1990 Network data and measurement. *Ann. Rev. Sociol.* **16**, 435–463. (doi:10.1146/annurev.so.16.080190.002251)

51. McCarty C. 2002 Measuring structure in personal networks. *J. Soc. Struct.* **3**, 1.

52. Jariego IM. 2018 Why name generators with a fixed number of alters may be a pragmatic option for personal network analysis. *Am. J. Community Psychol*. **62**, 233–238. (doi:10.1002/ajcp.12271)

53. Killworth PD, Johnsen EC, Bernard HR, Shelley AG, McCarty C. 1990 Estimating the size of personal networks. *Soc. Networks* **12**, 289–312. (doi:10.1016/0378-8733(90)90012-X)

54. McCarty C, Killworth PD, Rennell J. 2007 Impact of methods for reducing respondent burden on personal network structural measures. *Soc Networks* **29**, 300–315. (doi:10.1016/j.socnet.2006.12.005)

55. Stadel M, Stulp G. 2022 Balancing bias and burden in personal network studies. *Soc. Networks.* **70**, 16–24. (doi:10.1016/j.socnet.2021.10.007)

56. Dell'Era C, Landoni P. 2014 Living lab: A methodology between user-centred design and participatory design. *Creativ. Innov. Manag.* **23**, 137–154. (doi:10.1111/caim.12061)

57. Ruijsink S, Smith A. 2016 European Network of Living Labs (ENoLL) [Internet]. [cited 2024 May 16]. Available from: http://www.transitsocialinnovation.eu/resource-hub/european-network-of-living-labs

58. Hâncean M-G, Lubbers MJ, Molina JL 2021 Measuring transnational social fields through binational link-tracing sampling. *PLoS One* **16**, e0253042. (doi:10.1371/journal.pone.0253042)

59. Heckathorn DD. 1997 Respondent-driven sampling: a new approach to the study of hidden populations. *Soc. Probl.* **44**, 174–199. (doi:10.2307/3096941)

60. Merli MG, Verdery A, Mouw T, Li J. 2016 Sampling migrants from their social networks: the demography and social organization of Chinese migrants in Dar es Salaam, Tanzania. *Migr. Stud.* **4**, 182–214. (doi:10.1093/migration/mnw004)

61. Institutul Național de Statistică. 2021 Populația rezidentă pe sexe și grupe de vârstă [Internet]. [cited 2024 July 16]. Available from: https://insse.ro/cms/sites/default/files/com_presa/com_pdf/poprez_ian2023r.pdf

62. Complex Data Collective. 2023 Network Canvas: Interviewer (v6.5.2) [Computer software]. Zenodo. (doi:10.5281/zenodo.6026548)

63. McCarty C, Bernard HR, Killworth PD, Shelley GA, Johnsen EC. 1997 Eliciting representative samples of personal networks. *Soc Networks* **19**, 303–323. (doi:10.1016/s0378-8733(96)00302-4)

64. Hâncean M-G, Lerner J, Perc M, Molina JL, Geantă M. 2024a Assortative mixing of opinions about COVID-19 vaccination in personal networks. *Sci. Rep.* **14**, 3385. (doi:10.1038/s41598-024-53825-3)

65. Hâncean M-G, Lerner J, Perc M, Molina JL, Geantă M, Oană I, et al. 2024b Processed food intake assortativity in the personal networks of East European older adults [Internet]. medRxiv [Preprint] [cited 2024 Jun 29]. (doi:10.1101/2024.01.25.24301787)

66. Cepeda-Benito A, Doogan NJ, Redner R, Roberts ME, Kurti AN, Villanti AC, et al. 2018 Trend differences in men and women in rural and urban U.S. settings. *Prev. Med.* **117**, 69–75. (doi:10.1016/j.ypmed.2018.04.008)

67. Hunt LJ, Covinsky KE, Cenzer I, Espejo E, Boscardin WJ, Leutwyler H, et al. 2023 The epidemiology of smoking in older adults: a national cohort study. *J. Gen. Intern. Med.* **38**, 1697–1704. (doi:10.1007/s11606-022-07980-w)

68. Joung MJ, Han MA, Park J, Ryu SY. 2016 Association between family and friend smoking status and adolescent smoking behavior and E-cigarette use in Korea. *Int. J. Environ. Res. Public Health* **13**, 1183. (doi:10.3390/ijerph13121183)

69. Lastunen A, Laatikainen T, Isoaho H, Lazutkina G, Tossavainen K. 2017 Family members' and best friend's smoking influence on adolescent smoking differs between Eastern Finland and Russian Karelia. *Scand J Public Health* **45**, 789–798. (doi:10.1177/1403494817723550)

70. Higgins ST, Kurti AN, Redner R, White TJ, Gaalema DE, Roberts ME, et al. 2015 A literature review on prevalence of gender differences and intersections with other vulnerabilities to tobacco use in the United States, 2004-2014. *Prev. Med.* **80**, 89–100. (doi:10.1016/j.ypmed.2015.06.009)

71. Henley SJ, Asman K, Momin B, Gallaway MS, Culp MBB, Ragan KR, et al. 2019 Smoking cessation behaviors among older U.S. adults. *Prev. Med. Rep.* **16**, 100978. (doi:10.1016/j.pmedr.2019.100978)

72. Sachs-Ericsson N, Schmidt NB, Zvolensky MJ, Mitchell M, Collins N, Blazer DG. 2009 Smoking cessation behavior in older adults by race and gender: the role of health problems and psychological distress. *Nicotine Tob. Res.* **11**, 433–443. (doi:10.1093/ntr/ntp002)

73. Bauman KP, Geher G. 2002 We think you agree: The detrimental impact of the false consensus effect on behavior. *Curr. Psychol.* **21**, 293–318. (doi:10.1007/s12144-002-1020-0)

74. Ross L, Greene D, House P. 1977 The "false consensus effect": an egocentric bias in social perception and attribution processes. *J. Pers. Soc. Psychol.* **13**, 279–301. (doi:10.1016/0022-1031(77)90049-x)

75. Brewer DD, 2000 Forgetting in the recall-based elicitation of personal and social networks. *Soc. Networks* **22**, 29–43. (doi:10.1016/S0378-8733(99)00017-9)

76. Bergen N, Labonté R. 2019 "Everything is perfect, and we have no problems": detecting and limiting social desirability bias in qualitative research. *Qual. Health. Res.* **30**, 783–792. (doi:10.1177/1049732319889354)

77. Latkin CA, Edwards C, Davey-Rothwell MA, Tobin KE. 2017 The relationship between social desirability bias and self-reports of health, substance use, and social network factors among urban substance users in Baltimore, Maryland. *Addict Behav*. **73**, 133–136. (doi:10.1016/j.addbeh.2017.05.005)

78. Mihăilă B-E, Hâncean M-G, Perc M, Lerner J, Oană I, Geantă M, et al. 2024 Data from: Cross-sectional personal network analysis of adult smoking in rural areas [Dataset]. Zenodo. https://doi.org/10.5281/zenodo.13374383 (doi: 10.5281/zenodo.13374382)